\def\XXint#1#2#3{{\setbox0=\hbox{$#1{#2#3}{\int}$}
     \vcenter{\hbox{$#2#3$}}\kern-.5\wd0}}

\documentclass[aps,prb,twocolumn,showpacs,floatfix,eqsecnum]{revtex4}

\pdfoutput=1

\usepackage{times}
\usepackage{amsfonts}
\usepackage{amssymb}
\usepackage{amsmath}
\usepackage{graphicx}
\usepackage{bm}
\usepackage{verbatim}

\usepackage{hyperref}

\usepackage{bm} 
\usepackage{color}
\usepackage{stackrel}
\usepackage{accents}

\usepackage{latexsym}

\newcommand{\bsub}{\begin{subequations}}
\newcommand{\esub}{\end{subequations}}

\newcommand \bea {\begin{eqnarray} }
\newcommand \eea {\end{eqnarray}}
 
\newcommand{\beg}{\begin{equation}}
\newcommand{\en}{\end{equation}}

\newcommand{\bk}{\mathbf k}

\newcommand \bel  {\begin{align}}
\newcommand \enl  {\end{align}}

\newcommand{\up}{\uparrow}
\newcommand{\dn}{\downarrow}

\newcommand{\kk}{\mathbf{k}}
\newcommand{\Aq}{\mathbf{A}}

\newcommand{\pmat}{\begin{pmatrix}}
\newcommand{\epmat}{\end{pmatrix}}

\def\8{\infty}

\def\undertext#1{\vtop{\hbox{#1}\kern 1pt \hrule}}

\def\be{\begin{equation}}
\def\ee{\end{equation}}
\def\bea{\begin{eqnarray} & &}
\def\eea{\end{eqnarray}}


\begin{document}

\title{Short-time dynamics in $s+is$-wave superconductor with incipient bands}

\author{Marvin A. M\"{u}ller$^{1}$, Pengtao Shen$^2$, Maxim Dzero$^2$ and Ilya Eremin$^1$}
\affiliation{$^1$ Institut f\"{u}r Theoretische Physik III, Ruhr-Universit\"{a}t Bochum, D-44801 Bochum, Germany \\ 
$^2$Department of Physics, Kent State University, Kent, OH 44240 USA}

\begin{abstract} 
Motivated by the recent observation of the time-reversal symmetry broken
state in K-doped BaFe$_2$As$_2$ superconducting alloys, we theoretically study the
collective modes and the short time dynamics of the superconducting state with $s+is$-wave order parameter using an
effective four-band model with two hole and two electron pockets.
The superconducting $s+is$ state emerges for incipient electron bands as a result of hole doping and appears as an intermediate state between $s^{\pm}$ (high number of holes) and $s^{++}$ (low number of holes).
The amplitude and phase modes are coupled giving rise to a variety of collective modes.
In the $s^{\pm}$ state, we find the Higgs mode at frequencies similar to a two-band model with an
absent Leggett mode, while in the $s+is$ and $s^{++}$ state, we uncover a new coupled collective soft mode. Finally we compare our results with the $s+id$ solution and find similar behaviour of the collective modes.
\end{abstract}

\pacs{05.30.Fk, 32.80.-t, 74.25.Gz}

\maketitle

\section{Introduction}

Ultrafast pump-probe experiments recently become a powerful tool to probe the temporal dynamics of symmetry broken states and relaxation in conventional and unconventional superconductors\cite{Pashkin2010,Beck2011,Cav2011Science,Matsunaga2012,Conte2012,Beck2013,Matsunaga2013,Mansart2013,Shimano2014Science,CavNature2014,CavPRB2015,Matsunaga2017}. For high frequency excitation with
frequency exceeding the superconducting gap, $\Delta$, the radiation 
breaks Cooper pairs into quasiparticles, which yields rapid dissipation and
thermalization of the system. However, an intense pulse as used in Ref.\cite{Matsunaga2013}
couples non-linearly to the Cooper pairs  of the superconductor. This, as was argued theoretically, should lead to a coherent excitation of the
Higgs amplitude mode $\Delta(t)$\cite{Volkov1974,Amin2004,Barankov2004,Yuzbashyan2005,Yuzbashyan2006,Barankov2006,Papenkort2007,Krull2014,Big-Quench-Review2015,Tsuji2015,Aoki-Higgs2,Chou2017}. In the experiments \cite{Matsunaga2012,Matsunaga2013,Matsunaga2017} the detection is performed over a window of about 10 picoseconds
(ps), well before thermalization occurs (likely due
to acoustic phonons on a timescale of 100 ps\cite{Demsar2003}). 

While non-equilibrium collective modes in conventional single-gap superconductors are relatively well understood, the investigation of collective excitations in unconventional
non-equilibrium superconductors with multicomponent or multiple gaps is a very intriguing topic due to a very rich spectrum of the collective excitations \cite{Akbari2013,Dzero2015,Demsar2003b,Leggett1966,Anishchanka2007,Krull2017}. Fe-based superconductors are particularly interesting in this regard due to its variety and complexity of their phase diagrams. For example, recent experimental studies of the Fe-based superconductors have demonstrated the emergence of superconducting state with incipient bands, i.e. bands which do not cross Fermi level. \cite{Shin2014,Ding2015} Furthermore, the formation of the incipient bands is often connected to the Lifshitz transition, where one of the bands
continuously moves away from the Fermi level as a function of doping \cite{Lifshitz1960}. A peculiar example of an iron-based superconductor with
a Lifshitz transition is Ba$_{1−x}$K$_x$Fe$_2$As$_2$ where the Fermi surface topology changes from the one having both the electron and the hole pockets to the one where the electron pockets sink below the Fermi level. Angle-resolved photoemission
(ARPES)\cite{Sato2009,Xu2013} and thermopower\cite{Hodovanets2014} measurements
point toward the existence of such a transition in the overdoped hole-doped compounds with $x \sim 0.7–-0.9$. Intriguingly, in the
same doping range, the structure of the superconducting gaps
undergoes dramatic changes, seemingly inconsistent with a
two-band description. While multiple experiments supports the nodeless $s^{+-}$-wave superconducting gap near the optimal doping $x\sim 0.4$\cite{Nakayama2011,Ding2008,Christianson2008,Luo2009}, the situation is very different for the extremely overdoped case. In this doping range the experiments indicate either strongly anisotropic $s-$wave Cooper-pairing with sign change on the remaining two hole pockets\cite{Okazaki2012,Watanabe2014,Cho2016} or the $d$-wave pairing with well-defined nodes\cite{Reid2012,Tafti2013} on the hole Fermi surface sheets. Moreover, in the intermediate
doping region, frustration between the two superconducting
channels has been theoretically predicted to result in a time-reversal
symmetry-breaking $s + is$ state\cite{Stanev2010,Carlstroem2011,Hu2012,Maiti2013,sisEremin2017} or $s + id$ state\cite{Platt2009,Thomale2011}. In these states, the phase difference $\phi$ between the
order parameters at the two hole bands is not equal to a multiple
of $\pi$ with the $\phi \to - \phi$ symmetry being spontaneously broken. 

The time-reversal symmetry breaking $s+is$ and $s+id$ states possess several interesting properties and should demonstrate an unusual dynamics. For example, as a result of simultaneous breaking of $U(1)$ and $Z_2$ symmetries, the vortex
fractionalization and unusual vortex cluster states have been
predicted to exist for the $s+is$ state\cite{Carlstroem2011}. The collective excitations of
the phase differences between order parameters of different
bands (Leggett modes) in the $s + is$ state is expected to have peculiar
phase-density nature\cite{Carlstroem2011} and have been predicted to soften
at the $s + is$ critical points\cite{Maiti2013,sisEremin2017}.
The time-reversal symmetry breaking is most directly manifested in spontaneous currents around
nonmagnetic impurities\cite{Maiti2015} or quench-induced domain walls\cite{Garaud2014}. The currents result in local magnetic fields in the superconducting
phase and provide a signature of the time-reversal symmetry broken state.
Recent results in Ba$_{1-x}$K$_x$Fe$_2$As$_2$ 
are consistent with $s + is$ state or $s+id$ state at x = 0.73\cite{Grinenko2017}, close to the region
where the Lifshitz transition is considered to occur. This observation stimulates to analyze further details of the time-reversal symmetry broken states in strongly overdoped iron-based superconductors.

Given these theoretical and experimental developments, in this paper we study the signatures of the $s+is$ Cooper-pairing in the pump-probe
spectroscopy. Specifically, we adopt the four-band model developed in Ref. \onlinecite{sisEremin2017} to analyze the pairing dynamics. In addition, we also analyze the nature of the collective excitations. One of the our main findings are the emergence of the $s+is$ pairing when
initially the superconductor is in the $s^{\pm}$ state and existence of the sharp collective in the $s+is$ state. In addition, we also studied the pump-probe dynamics in the $s+id$ state and find that this state shows a very similar soft mode dynamics as it is the case for the $s+is$ state.

Our paper is organized as follows. Sections II and III contain a description of the model and its ground state within the mean-field theory approximation. In Section IV we study the non-adiabatic dynamics of the of a $s^{\pm}$ superconductor initiated by a sudden change of the pairing strength or an application of an external electromagnetic field. In Section V we present our results for the collective excitations depending on the doping level. Section VI is devoted to the discussion of our results.  

\section{Model}
We consider a model of a metal with four bands - two hole-like and two-electron like - with fully local particle-particle interactions. The model
Hamiltonian is: \cite{sisEremin2017}
\beg\label{Eq1}
\begin{split}
\hat{H}&=\sum\limits_{\bk,a,\sigma}\xi_\bk^a\hat{c}_{\bk\sigma}^{a\dagger}\hat{c}_{\bk\sigma}^a+\\&+
\sum\limits_{\bk\bk'}\sum\limits_{ab}\left(U_{ab}\hat{c}_{\bk\up}^{a\dagger}\hat{c}_{-\bk\dn}^{a\dagger}\hat{c}_{-\bk'\dn}^{b}\hat{c}_{\bk'\up}^b+
\textrm{h.c.}\right).
\end{split}
\en
Here $\{a,b\}\in\{h_1,h_2,e_1,e_2\}$ are the band labels, $\hat{c}_{\bk\sigma}^{a\dagger}$ ($\hat{c}_{\bk\sigma}^a$) are the fermionic creation (annihilation) operators, $U_{ab}>0$ are coupling constants and $\xi_\bk^a$ are the single particle dispersions in each band. In what follows, we
simplify our model by considering two identical electron- and hole-like bands within the tight-binding approximation:
\beg\label{disperse}
\begin{split}
&\xi_{\bk}^{h_i}=t_h(2-\cos k_x-\cos k_y)+\frac{E_D}{2}-\mu, \\
&\xi_{\bk}^{e_i}=t_e(2-\cos k_x-\cos k_y)-\frac{E_D}{2}-\mu,
\end{split}
\en
where $i=1,2$, $t_{e,h}$ are the hopping amplitudes, $\mu$ is the chemical potential and $E_D$ accounts for the changes in the relative occupation numbers of the electron and hole bands. 

Lastly, we set the coupling constants as 
\beg\label{Couplings}
\begin{split}
&U_{h_1h_1}=U_{h_2h_2}=0, \quad U_{e_1e_1}=U_{e_2e_2}=0, \\
&U_{h_1h_2}=U_{h_2h_1}=U_{\textrm{hh}}, \quad U_{e_ih_j}=U_{h_je_i}=U_{\textrm{eh}}
\end{split}
\en
(in the last expression $i\not=j$). Having formulated the model, we now review its ground state properties within the mean-field approximation.
\section{Mean-field theory}
Using the standard methodology, we decouple the two-fermion interaction term (\ref{Eq1}) in the particle-particle channel introducing
the following mean-field pairing amplitudes: $\Delta_e\propto\sum\limits_{\bk}\langle\hat{c}_{-\bk\dn}^{e}\hat{c}_{\bk\up}^e\rangle$ and
$\Delta_{h_{i}}\propto\sum\limits_{\bk}\langle\hat{c}_{-\bk\dn}^{h_{i}}\hat{c}_{\bk\up}^{h_{i}}\rangle$ ($i=1,2$). Minimizing the free energy with respect to the mean-field amplitudes yields the following system of equations
\beg\label{MFEqs}
\begin{split}
\Delta_e&=-U_{\textrm{eh}}\left(\Delta_{h_1}I_{h_1}+\Delta_{h_2}I_{h_2}\right), \\
\Delta_{h_1}&=-2U_{\textrm{eh}}\Delta_{e}I_e
-U_{\textrm{hh}}\Delta_{h_2}I_{h_2}, \\
\Delta_{h_2}&=-2U_{\textrm{eh}}\Delta_{e}I_e
-U_{\textrm{hh}}\Delta_{h_1}I_{h_1}.
\end{split}
\en
where we introduced
\beg\label{Is}
I_a=\sum\limits_{\bk}\frac{\tanh\left({E_{\bk}^{a}}/{2k_BT}\right)}{2E_{\bk}^{a}}, \quad a\in\{e,h_1,h_2\}
\en
for brevity and $E_{\bk}^a=\sqrt{(\xi_\bk^{a})^2+|\Delta_{a}|^2}$ are the single-particle energies. The mean-field equations (\ref{MFEqs}) have to be supplemented by the particle-number equation which determines the changes in the chemical potential due to the onset of superconducting order. It turns out to be convenient to evaluate the chemical 
potential as a function of the carrier number $n_c$ (per spin), i.e. the difference between the electrons and holes:
\beg\label{CarrierNumber}
\nonumber
\begin{split}
n_c&=\sum\limits_{\bk}\left[\frac{\xi_\bk^{e}}{E_{\bk}^{e}}\tanh\left(\frac{E_{\bk}^{e}}{2k_BT}\right)+\right.\\&\left.+\frac{\xi_\bk^{h}}{2E_{\bk}^{h_1}}\tanh\left(\frac{E_{\bk}^{h_1}}{2k_BT}\right)+\frac{\xi_\bk^{h}}{2E_{\bk}^{h_2}}\tanh\left(\frac{E_{\bk}^{h_2}}{2k_BT}\right)\right].
\end{split}
\en

\begin{figure}
\centering
\includegraphics[width=0.8\linewidth]{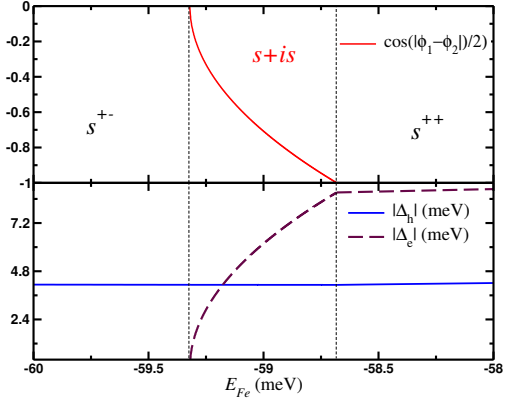}
\caption{Solution of the mean-field equations for the pairing amplitudes on the electron and hole pockets shown here as a function of the parameter $E_{Fe}=\mu+E_D/2$ with $E_D=0.12$ eV, $t_e=t_h=2.54$ eV, $U_{\textrm{hh}}=2.1925$ eV, $U_{\textrm{eh}}=2.3205$ eV. Note that the width of the $s+is$ region is approximately $1.5$ meV.}
\label{fig:Fig1}
\end{figure}

The detailed analysis of mean-field equations (\ref{MFEqs}) can be easily performed numerically. 
Upon closer inspection of these equations, however, it becomes
clear that one can basically guess the solution of these equations without resorting to numerics. Indeed, for the values of the chemical potential well below the bottom of the electronic bands, it is clear that there should be no pairing on the electronic pockets, $\Delta_e=0$, so from the mean field equations it follows that $\Delta_{h_1}=-\Delta_{h_2}$, while the $|\Delta_{h_1}|=|\Delta_{h_2}|=\Delta_h$ will be given by the root of $U_{\textrm{hh}}I_h[\Delta_{h}]=1$. Thus, for $\mu\ll -E_D/2$ the superconductivity is described by $s^{\pm}$ order parameter.
Let us now consider the opposite limit of filled electronic band, $\mu>-E_D/2$. Without any loss of generality, let us consider the pairing order parameter on the electron pockets to be purely real. Furthermore, the structure of the mean-field equations suggests that the pairing amplitudes in the hole bands must be the same:
\beg\label{assumptions}
\Delta_e=|\Delta_e|, \quad \Delta_{h_1}=|\Delta_h|e^{i\phi_1}, \quad \Delta_{h_2}=|\Delta_h|e^{i\phi_2}.
\en
We can now insert these equations into (\ref{MFEqs}) and separate the real and imaginary parts. It then follows that for the phases the following relation must hold
\beg\label{varphi}
\phi_1=-\phi_2=\varphi/2,
\en
where the relative phase $\varphi$ is determined by 
\beg\label{Eq4varphi}
\cos\left(\frac{\varphi}{2}\right)=
-\frac{U_{\textrm{hh}}}{2U_{\textrm{eh}}}\cdot\frac{|\Delta_e|}{|\Delta_h|},
\en
while the amplitudes $|\Delta_e|$ and $|\Delta_h|$ are the roots of the following two equations: $U_{\textrm{hh}}I_h\left(\Delta_h\right)=1$ and
\beg\label{Eq4DLThDLTe}
|\Delta_e|\cdot\left[1-{2U_{\textrm{eh}}^2I_e\left(\Delta_e\right)I_h\left(\Delta_h\right)}\right]=0.
\en
Clearly, these equations have a root corresponding to the $s^{\pm}$ state $(\Delta_e=0,\varphi=\pi)$ and a conventional $s$-wave state 
$(\Delta_e\not=0,\varphi=2\pi)$. It is therefore natural to expect that there also should be a solution corresponding to the intermediate values of phase $\varphi\in(\pi,2\pi)$. By analyzing the mean-field equations above numerically, we have confirmed that it is indeed the case and, 
moreover, the solution corresponding to the state with 
$\varphi\in(\pi,2\pi)$ has the lowest energy. In Fig. \ref{fig:Fig1} we present the results of the numerical analysis of the mean-field equations. 
Having reviewed the mean-field results for the model, we turn our discussion to the analysis of the collective response of the $s+is$ superconductor. 

\section{Temporal evolution of the $s+is$ order parameter}
In this Section we discuss the short-time dynamics of the $s+is$ superconductor initiated by either sudden change of the pairing strength or by a short pulse of an external electromagnetic field. Although at first glance these two ways of driving a system out-of-equilibrium seem to be very different, one can employ the linear analysis of the equations of motion, i.e. consider the limit of weak deviations from the ground state, to demonstrate that for the time-dependent correction to the ground state pairing amplitude external pulse has is many ways the same effect as a quench of the pairing strength. \cite{Aoki-Higgs1} 

To study the short time dynamics of a superconductor within the mean-field approximation described above, it is convenient to use the Anderson pseudospin variables. \cite{Anderson1958} In terms of these variables, the equations of motion are
\beg\label{EqOfMot}
\partial_t{\mathbf S}_{\bk}^l={\mathbf B}_\bk^l\times{\mathbf S}_\bk^l,
\en
where $l=\{h_1,h_2,e\}$ and $B_{\bk}^l=2(-\Delta_{lx},-\Delta_{ly},\xi_{\bk}^l)$. The $x$- and $y$-component of $B_{\bk}^l$ are given 
by the mean-field self-consistency equations
\beg\label{SelfConsSpins}
\begin{split}
\Delta_{h_1}^{\pm}(t)&=\sum\limits_{\bk}\left[-2U_{\textrm{eh}}S_{\bk e}^{\pm}(t)-U_{\textrm{hh}}S_{\bk h_2}^{\pm}(t)\right], \\
\Delta_{h_2}^{\pm}(t)&=\sum\limits_{\bk}\left[-2U_{\textrm{eh}}S_{\bk e}^{\pm}(t)-U_{\textrm{hh}}S_{\bk h_1}^{\pm}(t)\right], \\
\Delta_{e}^{\pm}(t)&=-U_{\textrm{eh}}\sum\limits_{\bk}\left[S_{\bk h_1}^{\pm}(t)+S_{\bk h_2}^{\pm}(t)\right]
\end{split}
\en
and we use the shorthand notation $\Delta^{\pm}=\Delta^x\pm i\Delta^y$. In the equilibrium a given pseudospin ${\mathbf S}_\bk^l$ is collinear with the corresponding $B_{\bk}^l$, so the sudden change of one of the coupling constants entering into (\ref{SelfConsSpins}) 
brings a system far from its equilibrium state.\cite{Barankov2004,Big-Quench-Review2015} 

\begin{figure}
\centering
\includegraphics[width=0.8\linewidth]{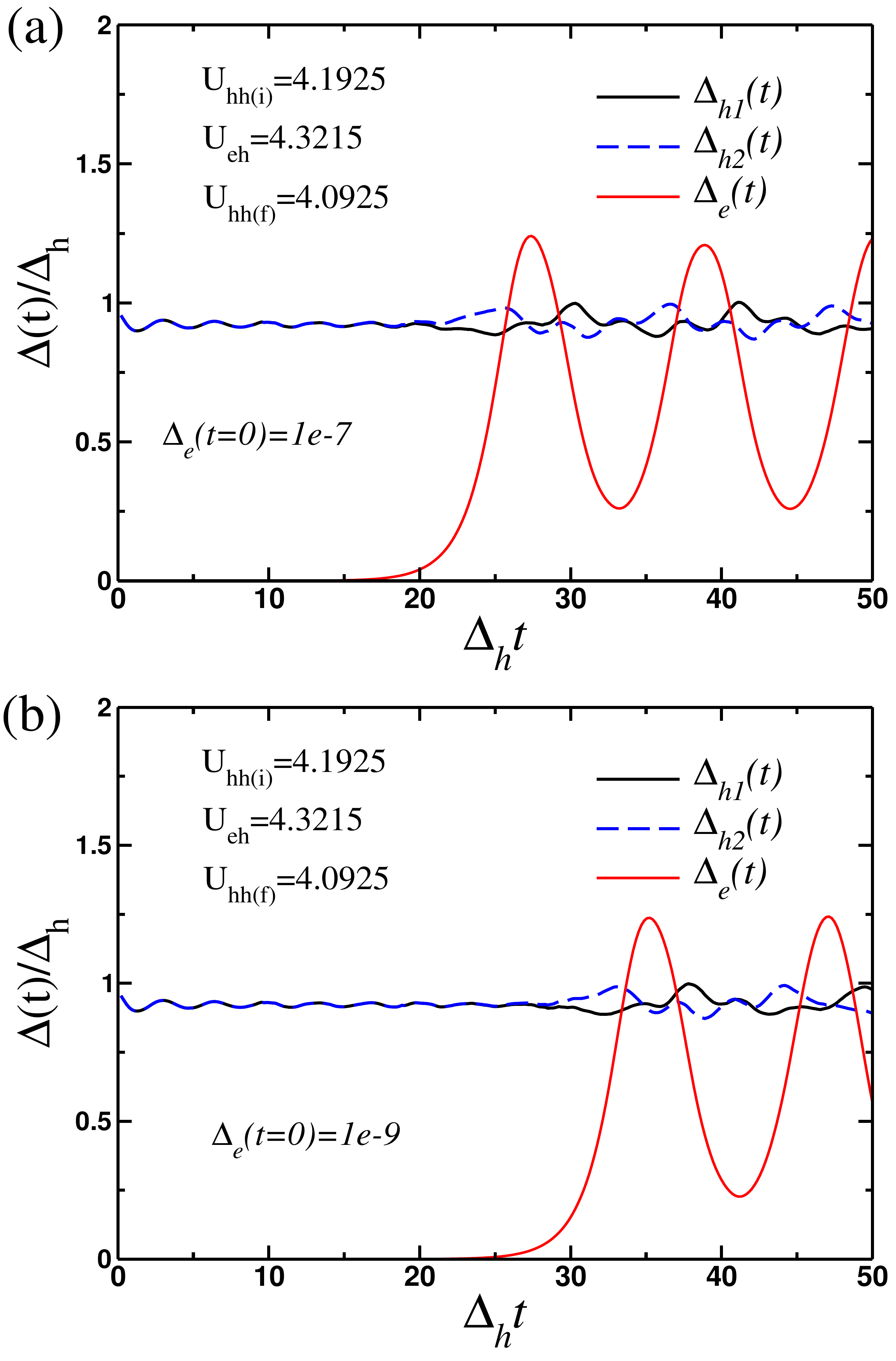}
\caption{Results of the numerical solution of the equations of motion (\ref{EqOfMot}) on the discreet momentum mesh with 
$N_{k_x}\times N_{k_y}=4096$ points. 
The values of the coupling parameters are chosen so that the initial state is $s^{\pm}$ near the boundary
separating the $s^\pm$ and $s+is$ ground states. Depending on the initial value of $\Delta_e$ [panels (a) and (b)] there is a critical time $t^{*}$ on which 
pairing amplitude on the electron pocket grows signaling the onset of the emergence $s+is$ state. }
\label{fig:Fig2}
\end{figure}

Given that the region in the parameter space of the relative band occupation numbers where the $s+is$ pairing state is a ground state is quite
narrow, we asked ourselves whether $s+is$ pairing amplitude will emerge dynamically when initially a superconductor is in the $s^\pm$ pairing ground state. To address this question, we solved the equations of motion numerically on a discreet mesh of momentum points. The results of the calculation are shown in Fig. \ref{fig:Fig2}. We indeed observe the dynamical onset of the $s+is$-pairing state, albeit this onset happens rather slowly. Specifically, it emerges on a time scale $t^*$ which far exceeds $\tau_{\Delta_e}\sim\hbar/\Delta_{e0}$ where $\Delta_{e0}$ is an equilibrium value corresponding to a ground state with the new coupling $U_{\textrm{hh}(f)}$.

In order to verify this result, we have performed the stability analysis by considering the pseudospin configuration corresponding to the $s^\pm$ state and allowing for small fluctuations into the $s+is$ state. Since numerical calculations show that $\delta\Delta_e(t)$ increases exponentially with time, we assume that
\beg\label{DlteExp}
\delta\Delta_e^{\pm}(t)=d_e^\pm e^{(\gamma\pm i\omega)t}, \quad d_e\ll |\Delta_h|
\en
and both $\gamma$ and $\omega$ are some real parameters which we need to determine. 
Employing the equations of motion, it is easy to show that the linear correction to the pseudospins on the electronic bands are
\beg\label{deltaSe}
\delta S_{\bk e}^{\pm}(t)=\pm\frac{\delta\Delta_e^{\pm}(t)}{i\gamma\mp\omega\pm 2\xi_\bk^e}.
\en
Similarly to (\ref{deltaSe}), the correction to the pseudospins on the hole bands are found to be
\beg
\begin{split}
&\delta S_{\bk h_1}^{\pm}(t)=\mp\frac{2\delta\Delta_{h_1}^{\pm}(t)S_{\bk z}^h}{i\gamma\mp\omega\pm 2\xi_\bk^h},\\
&\delta S_{\bk h_2}^{\pm}(t)=\mp\frac{2\delta\Delta_{h_2}^{\pm}(t)S_{\bk z}^h}{i\gamma\mp\omega\pm 2\xi_\bk^h}. 
\end{split}
\en
By inserting these expressions into  the self-consistency equations (\ref{SelfConsSpins}), one arrives at the system of linear equations for the 
linear corrections to the pairing fields on each pocket. These equations will have non-trivial solution provided the determinant of the corresponding matrix is zero. This condition has the form of the following equation
\beg\label{getgamma}
\left[1-U_{\textrm{hh}}\chi_h(\zeta)\right]\left[1+U_{\textrm{hh}}\chi_h(\zeta)-4U_{\textrm{eh}}^2\chi_e(\zeta)\chi_h(\zeta)\right]=0,
\en
where $\zeta=\gamma+i\omega$, $\chi_e(\zeta)=\sum_\bk(i\zeta+2\xi_\bk^e)^{-1}$ and $\chi_h(\zeta)=-\sum_\bk S_{\bk z}^h\cdot(i\zeta+2\xi_\bk^h)^{-1}$. The reader can easily check that this equation does not have a solution for $U_{\textrm{eh}}=0$ which 
means that the $s^{\pm}$ remains perfectly stable. However, for $U_{\textrm{eh}}\not =0$ such that the $s+is$ superconducting state is a ground state, we found that (\ref{getgamma}) has a root $\gamma\approx 10^{-3}\Delta_h$ and $\omega\ll 2\mu$.

\section{Collective modes}
\subsection{Response to fast perturbations: non-adiabatic regime}\label{sec_coll}
We now turn our discussion to the question of the collective response of the $s+is$ superconductor. Therefore, we perturb the system with a pump pulse to simulate the THz experiment. The electric field of this laser pulse can be described by a time dependent vector potential $\mathbf{E} = -\frac{1}{c}\partial_t\mathbf{A}$. We choose the spectrum of this pulse to be a Gaussian envelope centered around a frequency $\omega_0$ close to $\Delta_l$,
\begin{align}\label{eq:A}
\Aq(t) = \mathbf{A}_0\theta(-t)e^{-\frac{(t-t_0)^2}{2\tau^2}}\cos(\omega_0 t).
\end{align}
Here, $\tau$ controls the width of the time dependent signal and therefore needs to be chosen in such a way, that the signal disturbs the system non adiabatically. Also, we choose $\Aq_0$ small enough to not change the ground state, i.e., $\Delta_\infty^l \approx \Delta^l(0)$.
The electromagnetic vector potential couples to the system via minimal substitution, i.e., $\xi^l_\kk \rightarrow \xi^l_{\kk \pm \frac{e\Aq}{c}}$. This changes the pseudo magnetic field in the equations of motion (\ref{EqOfMot}) into
\begin{align}
\mathbf{B}_{\kk}^l = \left( -2\Delta_{lx} ,  -2\Delta_{ly}, \left(\xi_{\kk + \frac{e}{c}\Aq(t) l} + \xi_{\kk - \frac{e}{c}\Aq(t) l} \right) \right). 
\end{align}
\begin{figure}
	\centering
	\includegraphics[width=0.8\linewidth]{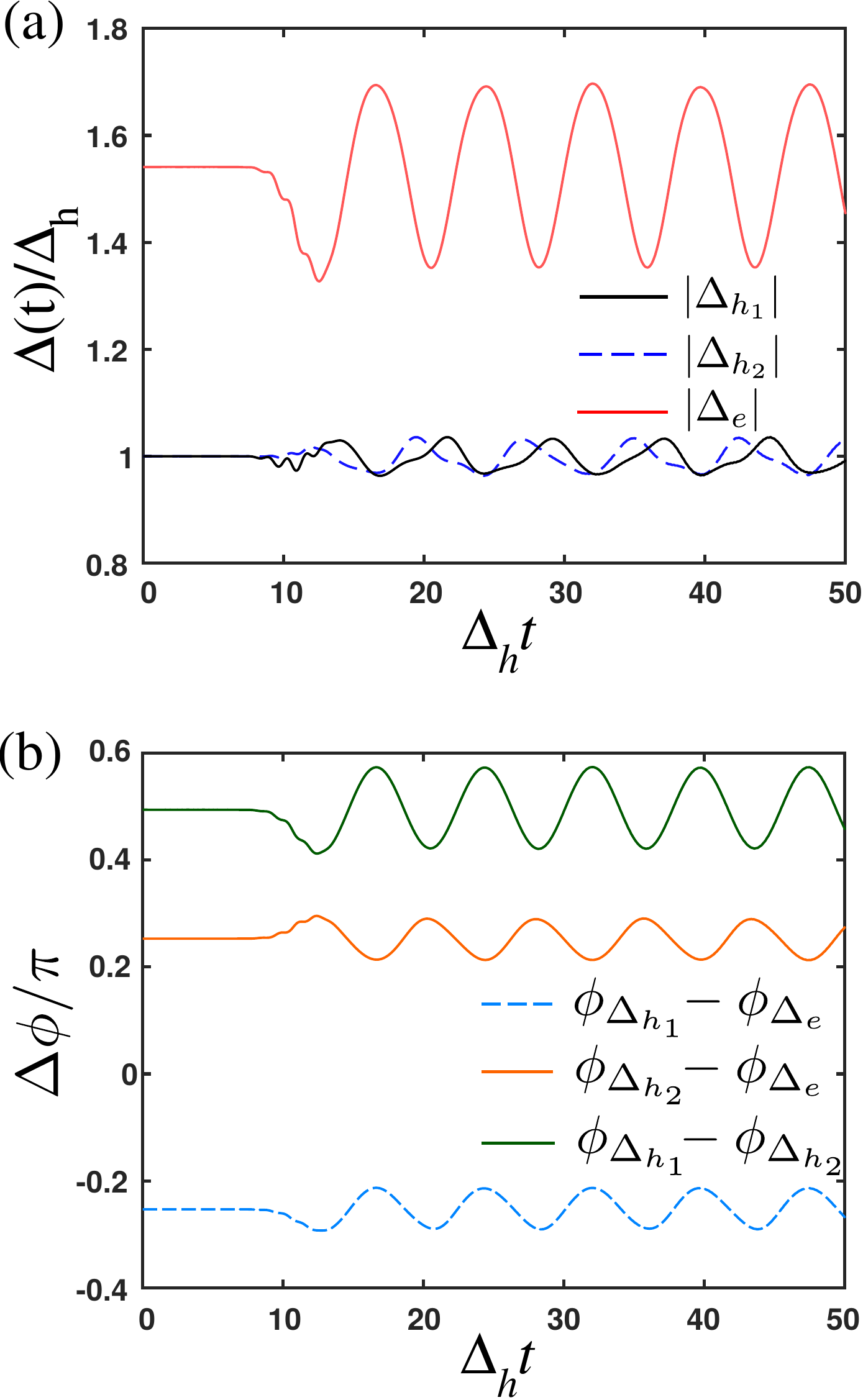}
	\caption{Numerical solution of the equations of motion with a time dependent vector potential $\Aq(t)$. The oscillation of both - the amplitude (panel (a)) and the relative phase (panel (b)) of the order parameters - are dominated by the same frequency $\omega_c$. }
	\label{fig:pulse}
\end{figure} \\
The numerical solution of the non-equilibrium dynamics are shown in Fig. \ref{fig:pulse}. One clearly see that both the dynamics of the amplitude and the relative phase are dominated by one frequency $\omega_c \sim 0.8 \Delta_{\infty}^h$. All oscillations are undamped, because the frequency is smaller than $2\Delta_{\infty}^h$, i.e., the smallest possible Higgs-mode and the beginning of the quasiparticle continuum. However, the absence of damping is a peculiarity of the model, since we assume isotropic $s$-wave order parameters. In the real system the nodal character of the order parameters leads to dampening effects.\\
\begin{figure}
	\centering
	\includegraphics[width=0.8\linewidth]{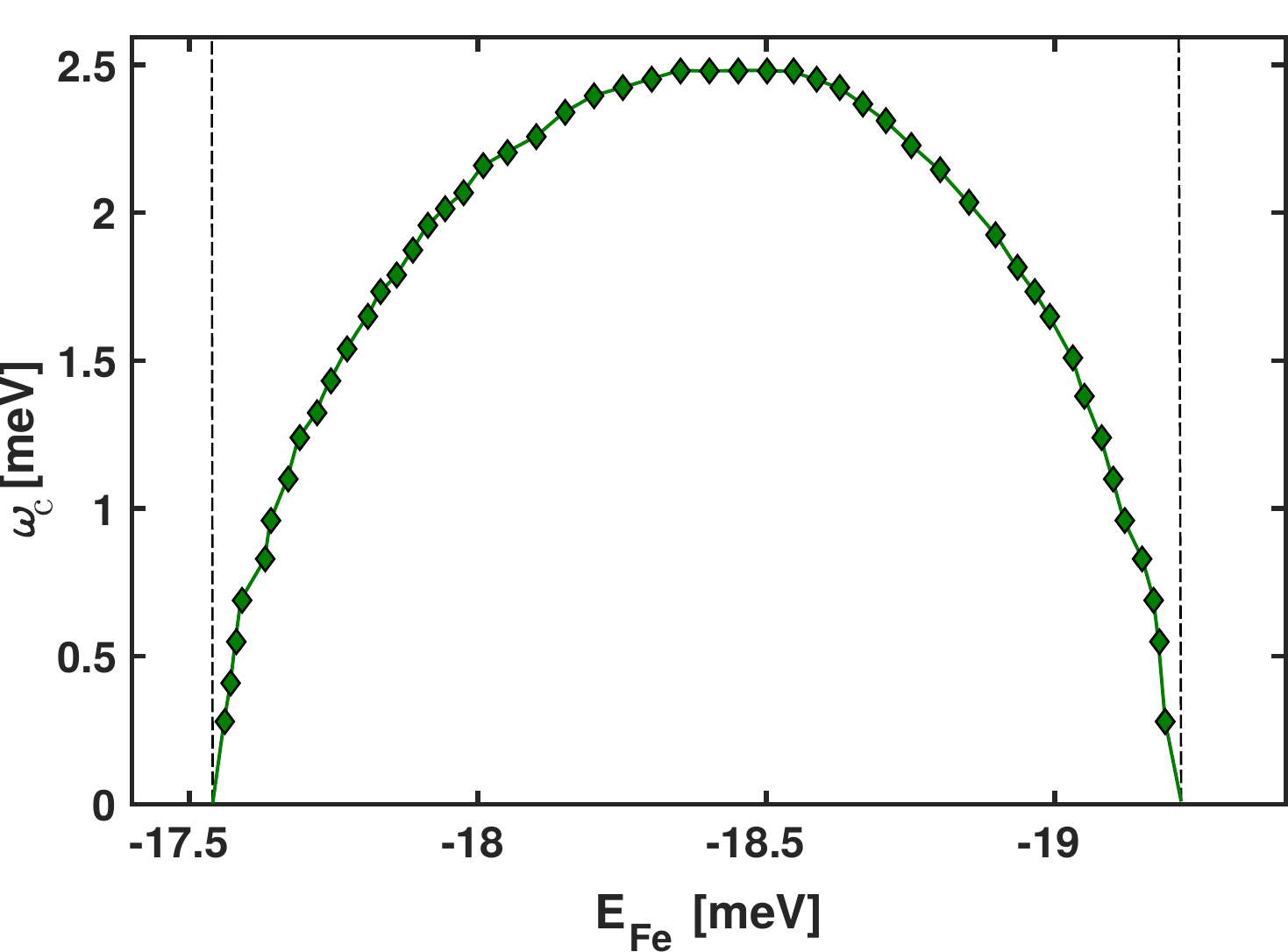}
	\caption{The mode $\omega_c$ across the full $s+is$ state. The dashed lines mark the borders of the TRSB state. For numerical accuracy we choose a cutoff energy ($\Lambda = 30$ meV) and $U_{hh} = 22$ meV and $U_{eh} = 20$ meV.}
	\label{fig:mode}
\end{figure} 
We examine the mode $\omega_c$ by numerically solving the equations of motions over the whole $s+is$ phase. The result is shown in Fig.~\ref{fig:mode}. Obviously, the mode $\omega_c$ softens close to the borders of the $s+is$-state and has its maximum value deep inside the $s+is$-regime. The coupling of both - amplitude and relative phase oscillations - is a peculiarity of the collective excitations in time reversal symmetry broken systems and differs from the usual dynamics of multiband superconductors. Also the softening of the mode close to the borders of the $s+is$ state can be explained by the second order phase transition from the $s+is$ into the $s^\pm / s^{++}$ state \cite{Carlstroem2011,Maiti2013, sisEremin2017}. For comparison we shown in Appendix In \ref{app_s+id} that this result still holds if the system, via an intermediate s+$i$d-state, ends up in a d-wave symmetry at large hole doping.
\subsection{Collective response in the adiabatic regime}
To study to collective mode at long wavelength limit, we linearize equations of motion (\ref{EqOfMot} ) with respect to the deviations from the equilibrium, $\delta {\mathbf S}_{\bk l}={\mathbf S}_{\bk l}(t)-{\mathbf S}_{\bk l}$, $\delta\Delta_{l}^x(t)=\Delta_{l}^x(t)-\Delta_{lx}$, $\delta\Delta_{l}^y(t)=\Delta_{l}^y(t)-\Delta_{l}^y$ and the effect of perturbation potential, $\delta B^z_{\bk l}(t)= B^z_{\bk l}(t)-\xi_{\kk}$. The deviations are homogeneous in space so they describe the collective mode at long wavelength limit. 
The details of the calculations can be found in the Appendix A. Here we discuss our main results. 

\begin{figure}[t]
	\centering
	\includegraphics[width=0.8\linewidth]{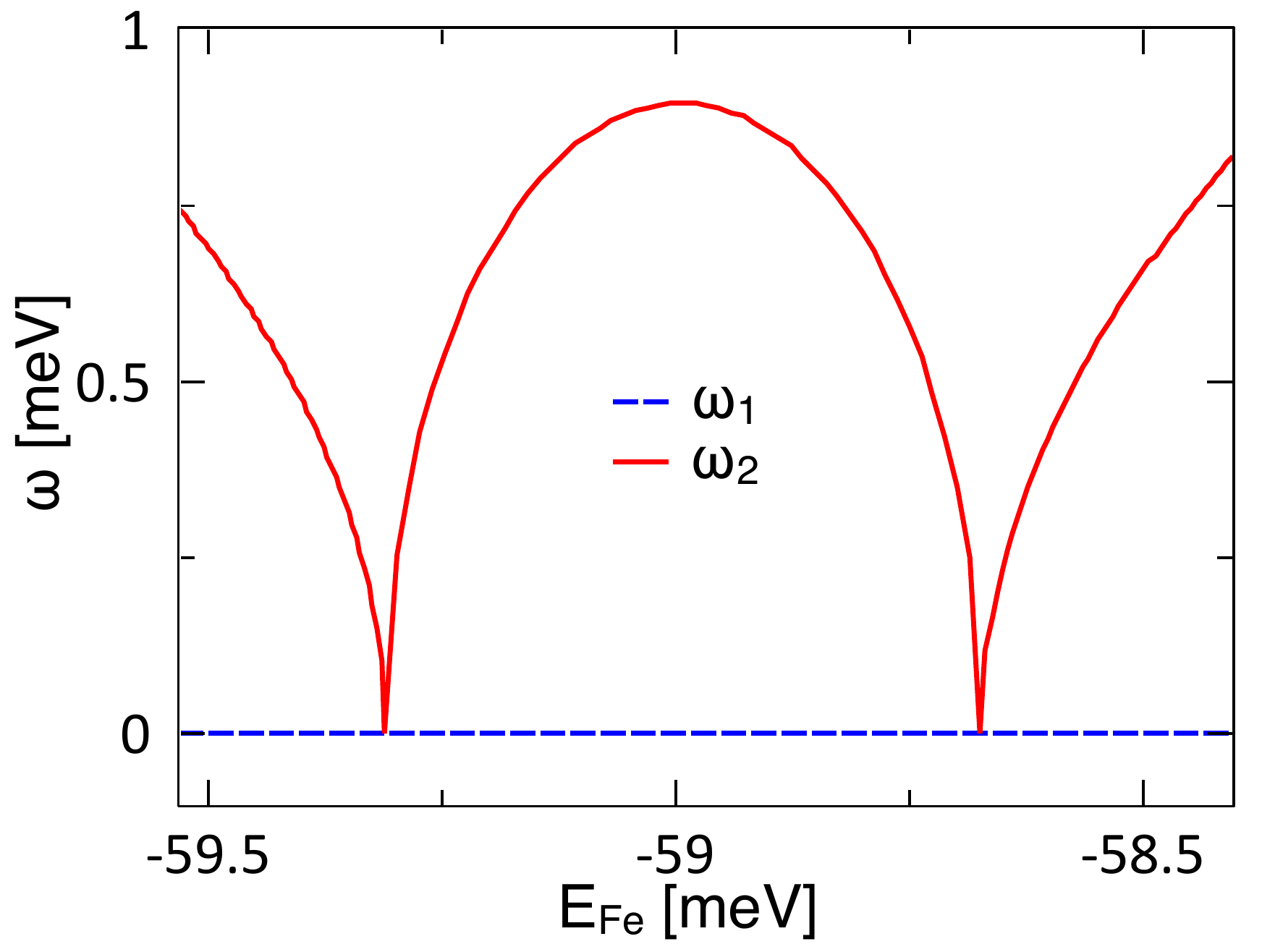}
	\caption{(color online) Mode  frequencies of collective mode at $q=0$ by varying doping parameter $E_{Fe}$: $\omega_1$ (blue dashed line) is the overall phase mode, $\omega_2$ (red solid line) is the coupled low energy mode. We choose $U_{\textrm{hh}}=2.1925$ eV and $U_{\textrm{eh}}=2.3205$ eV.}
	\label{fig:mode2}
\end{figure}

\begin{figure}[t]
	\begin{minipage}[t]{0.8\linewidth}
		\centering
		\includegraphics[width=0.7\linewidth]{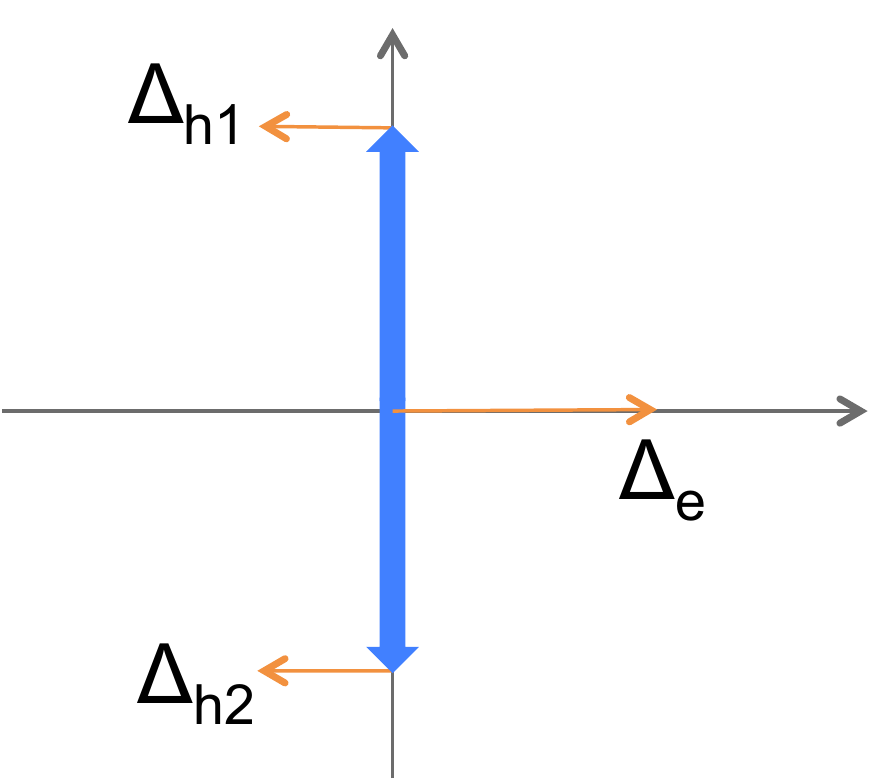}\\
	\end{minipage}
	\begin{minipage}[t]{0.8\linewidth}
		\centering
		\includegraphics[width=0.7\linewidth]{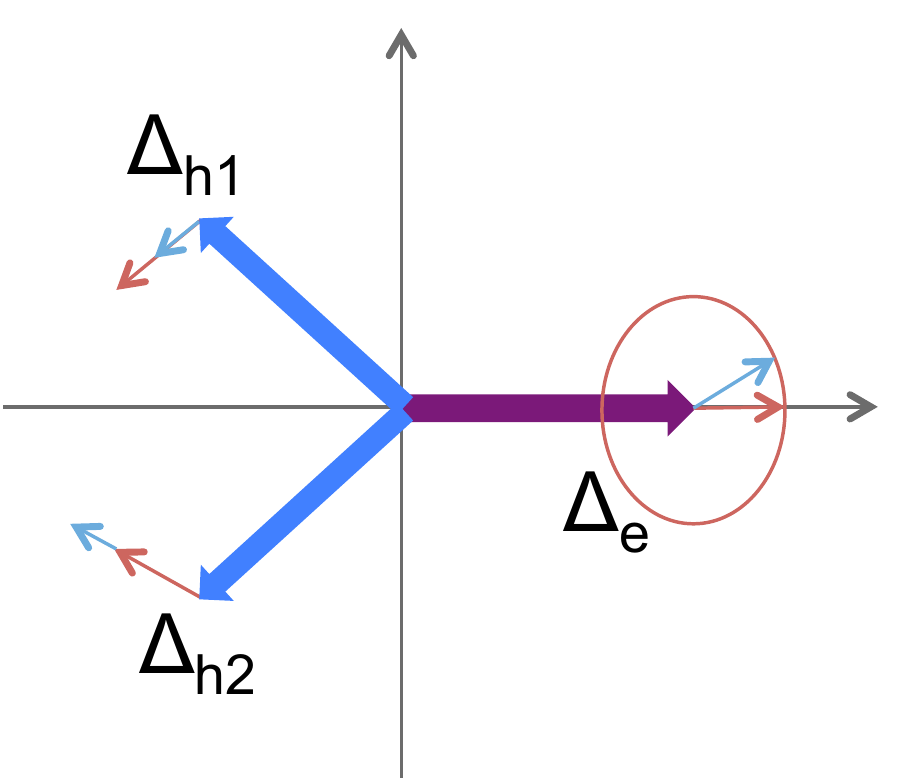}\\
	\end{minipage}
	\begin{minipage}[t]{0.8\linewidth}
		\centering
		\includegraphics[width=0.7\linewidth]{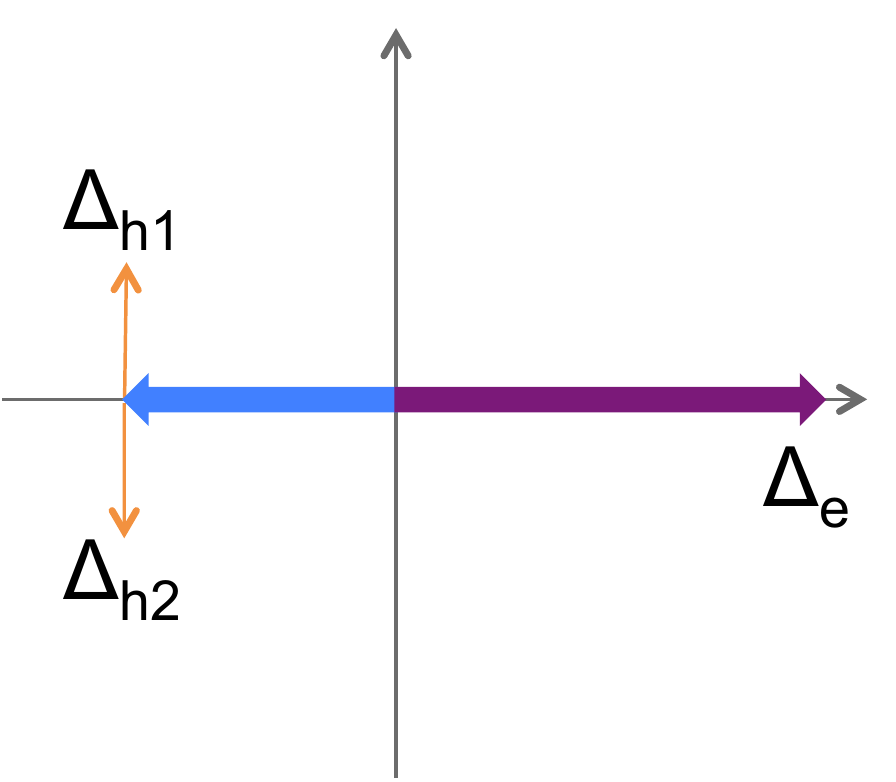}
	\end{minipage}
	\caption{Big arrows are the gap vectors in complex plane, small arrows are the mode vectors. From top to bottom are $s_{\pm}$, $s+is$ and $s_{++}$ state.}
	\label{fig:vector}
\end{figure}	

The resulting linearized equations will have non-trivial solution provided the corresponding determinant vanishes which sets us the non-linear equation for the frequency of the collective modes. 
We solve the equation for mode, Eq. (\ref{A3}), for all the ground states pseudospin configurations - $s^{\pm}$, $s+is$ and $s^{++}$ - and near the transitions between these states. In all three states, we find a solution at $\omega_1=0$: this mode is an overall phase mode without changing amplitude and relative phase difference, so this motion does not cost any energy. It is the Goldstone mode from U(1) symmetry breaking of BCS ground state.  Beside this, we also find a new low energy mode  $\omega_2$ as shown in Fig. \ref{fig:mode2}. The energy of this mode decreases at the boundary of $s+is$ state.

In $s^{\pm}$ state, we find that the mode is the motion of antisymmetric phase change of two hole bands gaps coupled with amplitude change of electron band gap. We have the eigenvector of mode,  $\delta{\vec \Delta}_{\pm}=[\delta\Delta_e,0,\delta\Delta_he^{-i\pi},0,\delta\Delta_he^{-i\pi},0]$, where  $\delta\Delta_e$, $\delta\Delta_h$ are positive real values, see Fig. \ref{fig:vector}. Similarly, in $s^{++}$ state, we find the mode is the motion of antisymmetric phase change of two hole bands gaps. We have the eigenvector of mode $\delta{\vec \Delta}_{++}=[0,0,\delta\Delta_he^{i\frac\pi2},0,\delta\Delta_he^{-i\frac\pi2},0]$

\begin{figure}[h]
	\centering
	\includegraphics[width=1\linewidth]{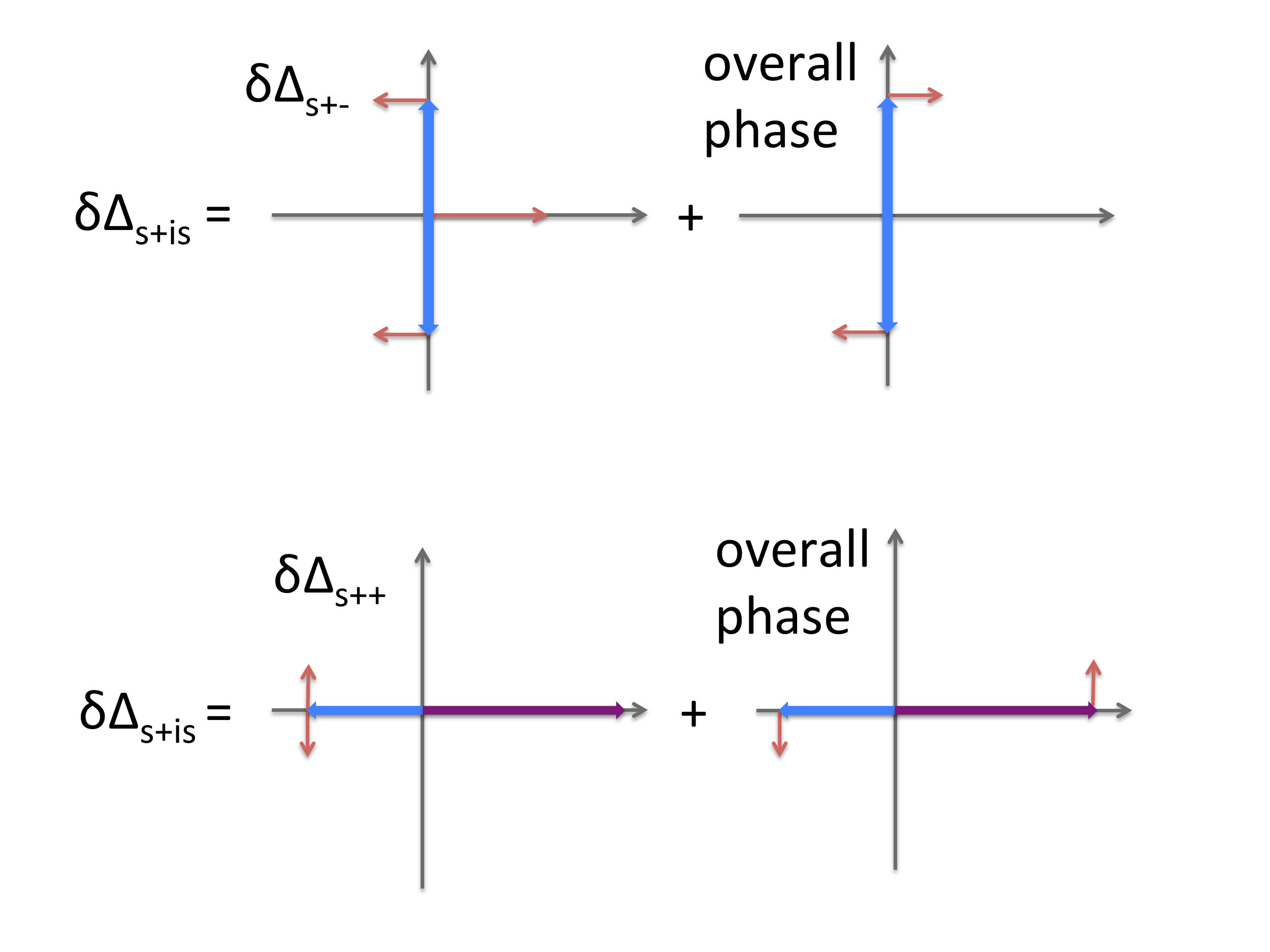}
	\caption{Decomposition of $s+is$ mode near the transition between $s^\pm$ and $s+is$ (top) and between  $s^{++}$ and $s+is$ (bottom). }
	\label{fig:decom}
\end{figure}

In $s+is$ state, the mode is an amplitude-phase coupled mode between both incipient electron and partially filled hole bands. Therefore it corresponds to $\omega_c$ in Section \ref{sec_coll}. The eigenvector is
\beg\label{Eigenvector}
\begin{split}
\delta{\vec \Delta}_{s+is}=[&\delta\Delta^x_e,\delta\Delta^y_e e^{\frac{i\pi}{2}},\delta\Delta^x_h e^{i\phi},\delta\Delta^y_h e^{i\phi},\\&\delta\Delta^x_he^{-i\phi},-\delta\Delta^y_h e^{-i\phi}]. 
\end{split}
\en
Note that the first two components of  $\delta{\vec \Delta}_{s+is}$ has a relative phase of $\pi/2$ which is a direct consequence of the incipiency of the electron band, as follows directly from the structure of the matrix elements (\ref{A3}) and the fact that $\Gamma_e(\omega)$ in that matrix is purely imaginary. Unlike the overall phase mode, this low energy mode vector is not continuous at the boundary between the two states. Near the boundary of $s^{\pm}$ and $s+is$ state, $\delta\Delta^y_e=\delta\Delta^y_h=0$, the mode is a combination of $\delta\vec\Delta_{\pm}$ mode and overall phase mode, see Fig. \ref{fig:decom}. At the end, it is not surprising because of both overall phase mode and low energy mode are soft at the boundary. As $E_{Fe}$ increasing, $\delta\Delta^x_e$, $\delta\Delta^x_h$ decrease and $\delta\Delta^y_e$, $\delta\Delta^y_h$ increase. Near the boundary of $s^{++}$ and $s+is$ state,  $\delta\Delta^x_e=\delta\Delta^x_h=0$, the mode is a combination of $\delta\vec\Delta_{++}$ mode and overall phase mode, Fig. \ref{fig:decom}.

\section{Discussion and Conclusion}
It is well known, the problem of non-adiabatic dynamics of the BCS model is exactly solvable \cite{Yuzbashyan2006}. 
The model presented above can be also shown to be exactly integrable for a special choice of the coupling constants, 
which however, do not describe the regime where $s+is$ state has the lowest energy. Nevertheless, we have checked that despite the lack of the integrability in our model, 
the non-equilibrium dynamics of the order parameter, which is initiated by a sudden change of the pairing strength 
bears a lot of similarities with the results obtained from exactly solvable version of the model. 

Another important aspect related to similarities and differences for the short-time order parameter dynamics between integrable and non-integrable models is concerned with the emergence of the $s+is$ state in the incipient electronic band. Indeed, within the both single-channel and two-channel pairing models for the degenerate atomic Fermi gases \cite{Big-Quench-Review2015} the realization of the steady state with periodically oscillating amplitude is limited to the weak-to-moderately strong coupling quenches in the vicinity of the BCS-BEC crossover. Our results show that, on one hand we have a steady state with periodically oscillating pairing amplitude, while on the other hand, the electronic band is incipient mimicking the BEC limit in atomic gases. Thus observation of this effect may question the interpretation of the 
$s+is$ state as being analogous to the BEC pairing in atomic condensates.

It is also important to keep in mind that the observation of the collective oscillations during the pump-probe experiments can in principle be inhibited by two effects: (i) spatial inhomogeneities of the pairing amplitude which may develop by parametric instabilities \cite{Dzero2008} and (ii) the Coulomb interactions between the particles. The effects of the Coulomb interactions on the non-equilibrium dynamics still remains a largely open problem. However, since our model involves purely repulsive interparticle interactions, we believe that the effects of the Coulomb interactions will not 
affect the dynamics in a profound way. 

To conclude, we theoretically study the
collective modes and the short time dynamics of the superconducting state with $s+is$-wave order parameter using an
effective four-band model with two hole and two electron pockets motivated by the recent experiments on time-reversal symmetry broken state in iron-based superconductors.
The superconducting $s+is$ state emerges for incipient electron bands as a result of hole doping and appears as an intermediate state between $s^{\pm}$ (high number of holes) and $s^{++}$ (low number of holes).
The amplitude and phase modes are coupled giving rise to a variety of collective modes.
In the $s^{\pm}$ state, we find the Higgs mode at frequencies similar to a two-band model with an
absent Leggett mode, while in the $s+is$ and $s^{++}$ state, we uncover a new coupled collective soft mode. We also compare our results with the $s+id$ solution and find similar behavior of the collective modes.

{\it Acknowledgments.} We are thankful to I. Paul and T. Tohyama for discussions. The work of P. S. and M. D. has been financially supported by the National Science Foundation grant NSF DMR-1506547. The work of M.D. was financially supported in part by the U.S. Department of Energy, Office of Science, Office of Basic Energy Sciences under Award No. DE-SC0016481. M.A.M. and I.E. were supported  by the joint DFG-ANR Project (ER 463/8-1) and DAAD PPP USA  N57316180.

\newpage

\appendix
\begin{widetext}
\section{collective mode in continuous model at q=0}
The linearized equations of motion are
\beg
\begin{split}
\partial_t\delta S^x_{\bk l}(t)=&-2\Delta^y_l\delta S^z_{\bk l}(t)-2 S^z_{\bk l}\delta\Delta^y_l(t)-2\xi_{\bk l}\delta S^y_{\bk l}(t)-2 S^y_{\bk l}\delta  B^z_{\bk l}(t),\\
\partial_t\delta S^y_{\bk l}(t)=&\quad 2\xi_{\bk l}\delta S^x_{\bk l}(t)+2 S^x_{\bk l}\delta  B^z_{\bk l}(t)+2\Delta^x_l\delta S^z_{\bk l}(t)+2 S^z_{\bk l}\delta\Delta^x_l(t),\\
\partial_t\delta S^z_{\bk l}(t)=&-2\Delta^x_l\delta S^y_{\bk l}(t)-2 S^y_{\bk l}\delta\Delta^x_l(t)+2 S^x_{\bk l}\delta\Delta^y_l(t)+2\Delta^y_l\delta S^x_{\bk l}(t) 
\end{split}
\en

Fourier transformation, e.g., $f(t)=\int \frac {d\omega}{2\pi}f(\omega)e^{i\omega t}$, give

\beg
\begin{split}
\left(
\begin{array}{c}
\delta S^x_{\bk l}(\omega)\\
\delta S^y_{\bk l}(\omega)\\
\delta S^z_{\bk l}(\omega)
\end{array}
\right)
&=\frac { S^z_{\bk l}}{\xi_{\bk l }(4E^2_{\bk l }-\omega^2)}
\left(
\begin{array}{ccc}
4(\xi^2_{\bk l }+{\Delta^y_l}^2) & 2i\omega\xi_{\bk l }-4\Delta^x_l\Delta^y_l& 2i\omega\Delta^y_l+4\Delta^x_l\xi_{\bk l }\\
-2i\omega\xi_{\bk l }-4\Delta^x_l\Delta^y_l & 4(\xi^2_{\bk l }+{\Delta^x_l}^2)& -2i\omega \Delta^x_l+4\Delta^y_l\xi_{\bk l }\\
-2i\omega\Delta^y_l+4\Delta^x_l\xi_{\bk l } & 2i\omega \Delta^x_l+4\Delta^y_l\xi_{\bk l }& 4({\Delta^x_l}^2+4{\Delta^y_l}^2)
\end{array}
\right)
\left(
\begin{array}{c}
-\delta\Delta^x_l(\omega)\\
-\delta\Delta^y_l(\omega)\\
\delta  B^z_{\bk l }(\omega)
\end{array}
\right)
\end{split}
\en
We take $\delta B^z_{\bk l}=0$ since the collective mode should not depend on the perturbation and substitute the result into  self-consistency equation (\ref{SelfConsSpins} ), we obtain
\beg\label{A3}
\left(
\begin{array}{cccccc}
1&0&U_{\textrm{eh}}M_{h1}^x(\omega)&U_{\textrm{eh}}\Gamma_{h1}(\omega)&U_{\textrm{eh}}M_{h2}^x(\omega)&U_{\textrm{eh}}\Gamma_{h2}(\omega)\\
0&1&U_{\textrm{eh}}\Gamma_{h1}(-\omega)&U_{\textrm{eh}}M_{h1}^y(\omega)&U_{\textrm{eh}}\Gamma_{h2}(-\omega)&U_{\textrm{eh}}M_{h2}^y(\omega)\\
2U_{\textrm{eh}}M_e^x(\omega)&2U_{\textrm{eh}}\Gamma_e(\omega)&1&0&U_{\textrm{hh}}M_{h2}^x(\omega)&U_{\textrm{hh}}\Gamma_{h2}(\omega)\\
2U_{\textrm{eh}}\Gamma_e(-\omega)&2U_{\textrm{eh}}M_e^x(\omega)&0&1&U_{\textrm{hh}}\Gamma_{h2}(-\omega)&U_{\textrm{hh}}M_{h2}^y(\omega)\\
2U_{\textrm{eh}}M_e^x(\omega)&2U_{\textrm{eh}}\Gamma_e(\omega)&U_{\textrm{hh}}M_{h1}^x(\omega)&U_{\textrm{hh}}\Gamma_{h1}(\omega)&1&0\\
2U_{\textrm{eh}}\Gamma_e(-\omega)&2U_{\textrm{eh}}M_e^x(\omega)&U_{\textrm{hh}}\Gamma_{h1}(-\omega)&U_{\textrm{hh}}M_{h1}^y(\omega)&0&1
\end{array}
\right)
\left(
\begin{array}{c}
\delta\Delta^x_e(\omega)\\
\delta\Delta^y_e(\omega)\\
\delta\Delta^x_{h_1}(\omega)\\
\delta\Delta^y_{h_1}(\omega)\\
\delta\Delta^x_{h_2}(\omega)\\
\delta\Delta^y_{h_2}(\omega)\\
\end{array}
\right)
=0
\en
where the quantities in the equation at T=0 are given by
\beg\label{A4}
\begin{split}
M_{l}^{\alpha}(\omega)=\sum_\bk \frac {2[\xi^2_{\bk l }+(\Delta_l^{\alpha})^2]}{(4E^2_{\bk l }-\omega^2)E_{\bk l }},\quad
\Gamma_l(\omega)=\sum_\bk \frac {i\omega\xi_{\bk l }-2\Delta^x_l\Delta^y_l}{(4E^2_{\bk l }-\omega^2)E_{\bk l }}
\end{split}
\en
and $\alpha=x,y$.

\end{widetext}

In usual one band or two bands superconductor, after choosing proper gauge, $\Gamma_l(w)=0$ due to electron-hole symmetry, thus the amplitude and phase are decoupled in the equation. In the time reversal symmetry broken system, one can not choose a gauge to vanish all $\Gamma_l(w)$. Besides, the broken electron-hole symmetry of the incipient electron band makes $\Gamma_e(w)$ always non-vanishing. As a result, the amplitude and phase oscillation are coupled in the mode. 
The frequency of the mode is determined by the solution of equation 
\beg\label{A5}
\textrm{Det}[\mathbf M]=0
\en
where $\mathbf M$ is the 6$\times$6 matrix in (\ref{A3}). The eigenvectors of the matrix 
\beg
\begin{split}
\delta{\vec \Delta}(\omega)= [&\delta\Delta^x_e(\omega),\delta\Delta^y_e(\omega),\delta\Delta^x_{h_1}(\omega),\\
&\delta\Delta^y_{h_1}(\omega),\delta\Delta^x_{h_2}(\omega),\delta\Delta^y_{h_2}(\omega) ]
\end{split}
\en
tell how amplitude and phase are coupled in the collective excitation. 

\section{collective modes inside the $s+id$ regime}\label{app_s+id}
\begin{figure}
	\includegraphics[width=\linewidth]{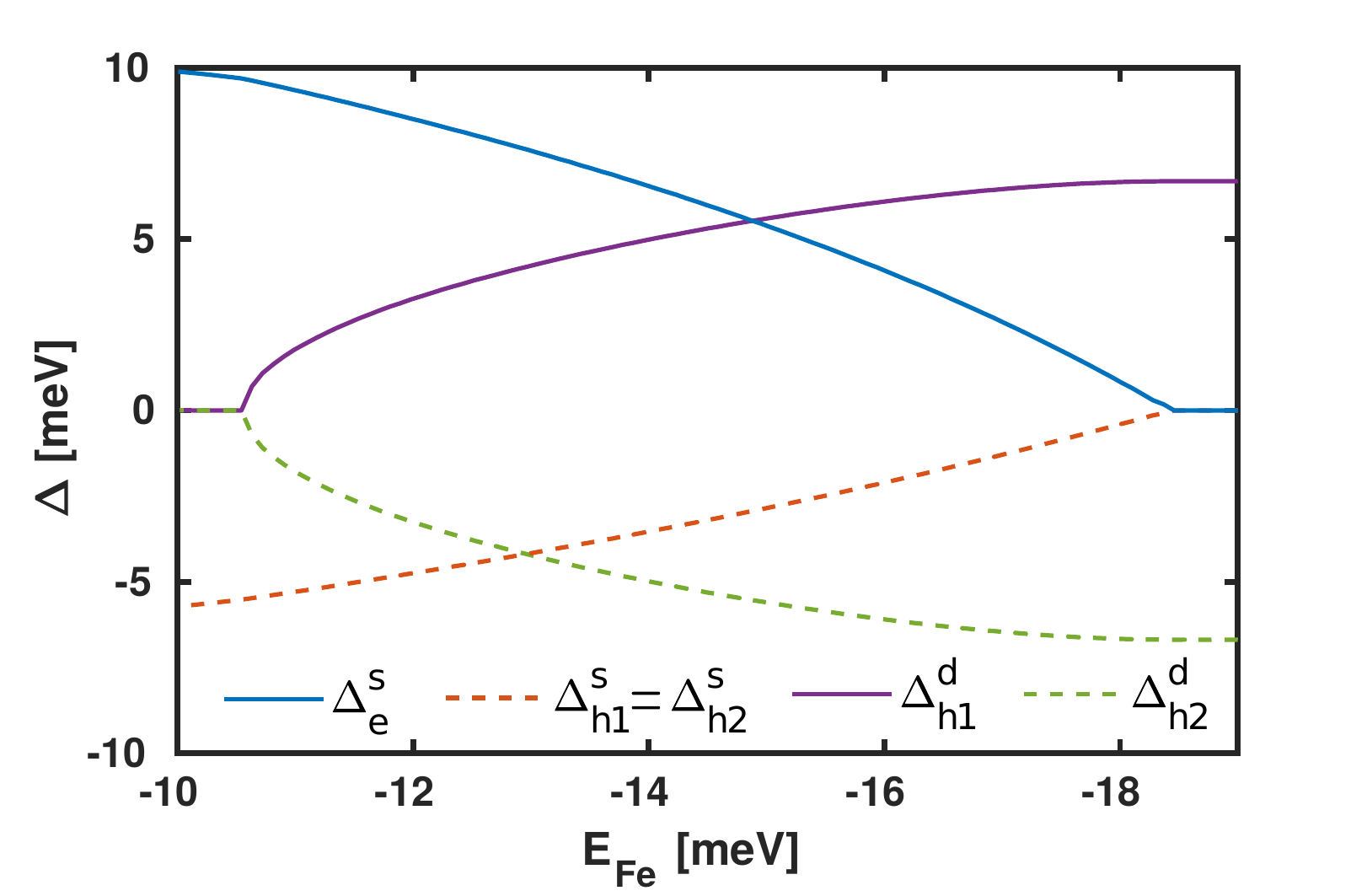}
	\caption{The solution of the mean-field equations at zero temperature. In this picture the different order parameters are plotted against $E_{Fe}$. Between $-10.5$ meV and $-18.5$ meV both hole order parameters consist out of a s-wave component and a d-wave component. We introduced an energy cutoff $\Lambda = 30$ meV and $U_{hh,s} = 20$ meV, $U_{hh,d} = 50$ meV and $U_{eh} = 22$ meV. \label{fig:s+id_phasediagram}}
\end{figure}

\begin{figure}[h]
	\includegraphics[width=\linewidth]{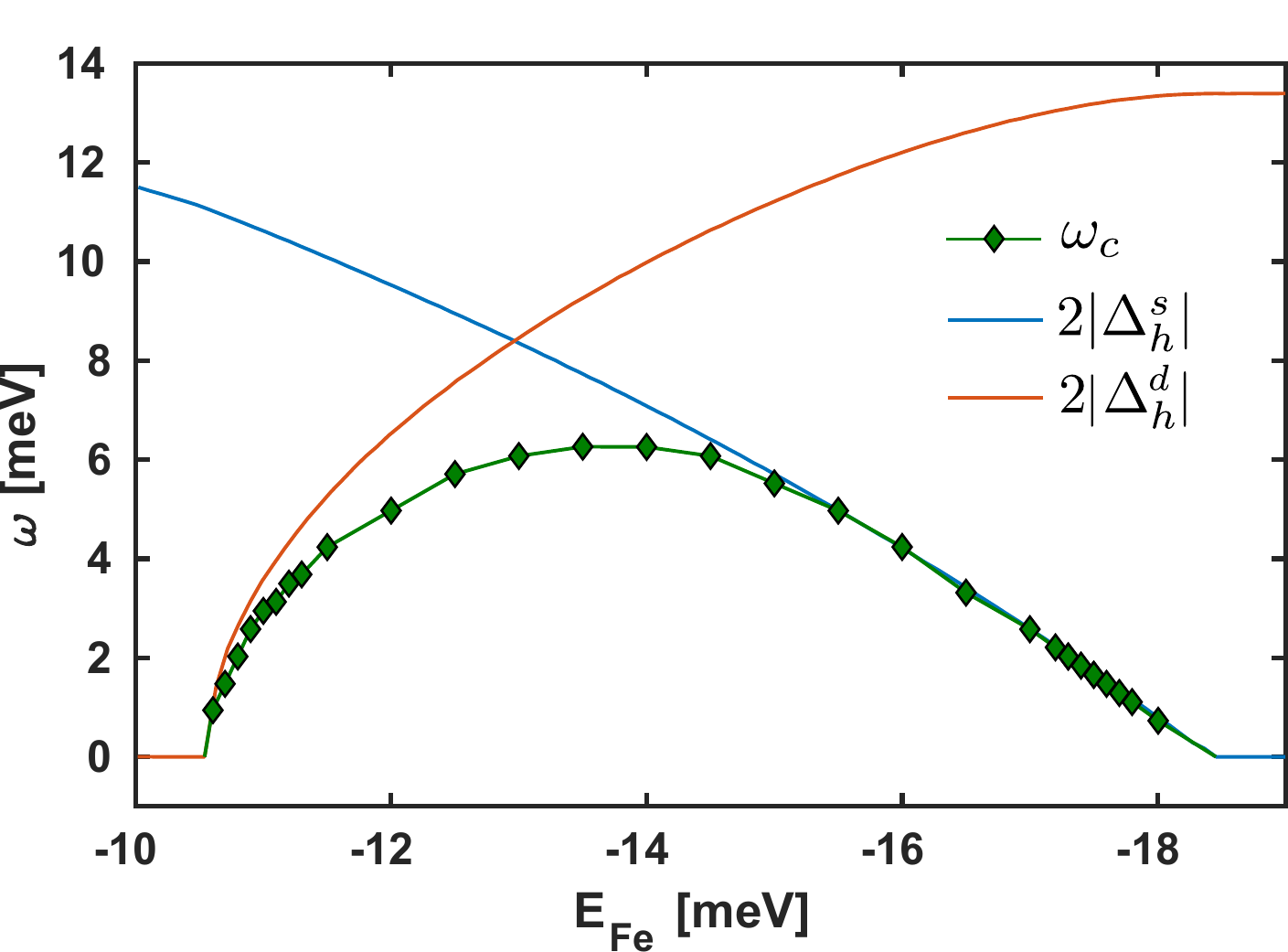}
	\caption{The frequency of the collective mode $\omega_c$ over the whole s+$i$d regime. Here we also compare the data to $2|\Delta_h^{s,d}| \equiv 2|\Delta_{h_1}^{s,d}| = 2|\Delta_{h_2}^{s,d}|$.  \label{fig:s+id_mode}}
\end{figure}

In this section we discuss the collective dynamics inside a s+$i$d superconductor. Therefore, we modify the Hamiltonian (\ref{Eq1}) and add another channel to the inter-hole-band interaction $U_{hh}$ to allow for d-wave pairing
\begin{align}
U^{h_1h_2}_{\kk,\kk^\prime} &\equiv U_{hh,s} + U_{hh,d}\cos(2\phi_\kk)\cos(2\phi_{\kk^\prime}),
\end{align}
where $U_{hh,s},U_{hh,d} > 0$ are constant and $\cos(2\phi_\kk) = (k_x^2 - k_y^2)/(k_x^2 + k_y^2)$. This leads to momentum dependent order parameters for the two hole bands
\begin{align}
\Delta_\kk^{h_i} = \Delta_{h_i}^s + i\Delta_{h_i}^d\cos(2\phi_\kk),
\end{align}
where $\Delta_{h_i}^s$ and $\Delta_{h_i}^d$ are the s- and d-wave pairing amplitude. It is important that the only possible pairing symmetry with mixed s- and d-wave component is s+$i$d, since s+d symmetry breaks the $C_4$-symmetry of the system. Also we choose the d-wave component large enough to suppress a possible competition between s+$i$s and s+$i$d superconductivity, i.e., we can choose $\Delta_{h_i}^s$ and $\Delta_{h_i}^d$ real. This argument can also be shown by free energy analysis. The pairing amplitude for the electron-band remains constant and is chosen positive to fix the overall phase. \\
Minimizing the free energy with respect to the five different pairing amplitudes we obtain
\begin{align}
\Delta_e &= -U_{eh}(\Delta_{h_1}^sI_{h_1} +\Delta_{h_2}^sI_{h_2}),\nonumber \\
\Delta_{h_1}^s &= -2U_{eh}\Delta_e I_e  -U_{hh,s}\Delta_{h_2}^sI_{h_2}, \nonumber\\
\Delta_{h_2}^s &= -2U_{eh}\Delta_e I_e  -U_{hh,s}\Delta_{h_1}^sI_{h_1},
\\
\Delta_{h_1}^d &= - U_{hh,d} \Delta_{h_2}^dJ_{h_2},\nonumber \\
\Delta_{h_2}^d &= - U_{hh,d} \Delta_{h_1}^dJ_{h_1}\nonumber,\end{align}
where $I_a$ was introduced in Eq. (\ref{Is}) and 
\begin{align}
J_a = \sum_\kk \frac{\tanh(E_{\kk}^a/2k_B T)}{2E_{\kk}^a}\cos^2(2\phi_\kk).
\end{align}
Since the order parameters are momentum dependent, the single-particle energies of the hole-bands now have the form $E_\kk^{h_i} = \sqrt{(\xi_\kk^{h_i})^2 + |\Delta_{h_i}^s|^2 + |\Delta_{h_i}^d|^2\cos^2(2\phi_\kk)}$. \\
Again, we take a closer look into this set of equations. Clearly, in the pure s-wave limit, i.e., $\Delta_{h_i}^d$ = 0, we reproduce our previous set of mean field equations in Eq. (\ref{MFEqs}). In the pure d-wave limit, i.e., $\Delta_{h_i}^s = 0$, $\Delta_e$ becomes zero and we end with only the last two equations and it follows $\Delta_{h_1}^d =-\Delta_{h_2}^d$. Obviously this scenario is described in the case $\mu\ll -E_D/2$, where the electron band is incipient and $\Delta_e = 0$. Since we choose $U_{hh,d}$ large enough to make the solution  $\Delta_{h_1}^s = -\Delta_{h_2}^s$ energetically unfavorable we end up with the pure d-wave solution. However for $\mu > -E_D/2$ we end up with finite $\Delta_e>0$ and thus finite $\Delta_{h_1}^s = \Delta_{h_2}^s <0$. For large enough $\Delta_e$, i.e., large enough $\mu$, it is energetically unfavorable for the system to condense an additional d-wave component and thus $\Delta_{h_1}^d =\Delta_{h_2}^d = 0 $. However, in between these two configurations we can end up in a state, where all five pairing amplitudes are finite. This set of equations is solved numerically in Fig. (\ref{fig:s+id_phasediagram}). 
 \\
Similar to Section IV we obtain the equations of motion for this model by making use of Anderson pseudospin variables. The equations of motions have the same form as in Eq. (\ref{EqOfMot}) but the pseudo magnetic field is now $B_\kk^l = 2(-\Delta_{\kk l}^x, -\Delta_{\kk l }^y, \xi_{\kk}^l)$, where we use $\Delta_{\kk l} = \Delta_l^s + i\Delta_{l}^d\cos(2\phi_{\kk})$ and $\Delta_{\kk l}^x, \Delta_{\kk l }^y$ as introduced in the main text. Here, one needs to keep in mind that the electron order parameter has only the s-wave component, i.e., $\Delta_{e}^d = 0$. The $x$- and $y$- component of the pseudo magnetic field are given by
\begin{align}
\Delta_{h_1}^{s,\pm}(t) =& -\sum_\kk \left[2U_{eh}S_{\kk e}^\pm(t) + U_{hh,s}S_{\kk h_2}^\pm(t)\right] \nonumber\\\nonumber
\Delta_{h_1}^{d,\pm}(t)=& -\sum_\kk U_{hh,d}S_{\kk h_2}^\pm(t)\cos(2\phi_\kk) \\
\Delta_{h_2}^{s,\pm}(t) =& -\sum_\kk \left[2U_{eh}S_{\kk e}^\pm(t) + U_{hh,s}S_{\kk h_1}^\pm(t)\right] \\\nonumber
\Delta_{h_2}^{d,\pm}(t)=& -\sum_\kk U_{hh,d}S_{\kk h_1}^\pm(t)\cos(2\phi_\kk) \\\nonumber
\Delta_e^\pm(t) =& -\sum_\kk\left[U_{eh}S_{\kk h_1}^\pm(t) + U_{eh}S_{\kk h_2}^\pm(t)\right].
\end{align}
Including a time-dependent vector potential $\Aq(t)$ changes the pseudo magnetic field in the equations of motion into
\begin{align}
\mathbf{B}_{\kk}^l = \left( -2\Delta_{\kk l}^x ,  -2\Delta_{\kk l}^y, \left(\xi_{\kk + \frac{e}{c}\Aq(t) l} + \xi_{\kk - \frac{e}{c}\Aq(t) l} \right) \right). 
\end{align}
Choosing $\Aq(t)$ as in Eq. (\ref{eq:A}) we solve the equations of motion for a system inside the s+$i$d regime numerically. Due to the momentum dependence of the order parameters the result is depending on the polarization of the vector potential. However, this does not change the qualitative dynamics. Similar to the s+$i$s scenario we obtain that the dynamics of both - amplitude and relative phase oscillation - are clearly dominated by a single frequency $\omega_c$. We find that all pairing amplitudes oscillate at the same frequency. While the relative phase between the hole order parameters remains constant for both s- and d-wave component, the relative phase between s- and d-wave component oscillate for each hole order parameter. \\
In Fig. (\ref{fig:s+id_mode}) we investigate this frequency over the whole s+$i$d phase diagram. We find that $\omega_c$ behaves similarly to the s+$i$s scenario and vanished close to the borders of the s+$i$d state. However, close to the border to the d-wave state one obtains that $\omega_c$ is coupling to $2|\Delta_{h_2}^{s}|$, which can be understood as the system's decreasing Higgs-mode due to the transition into the nodal d-wave state. This non-equilibrium effect is similar to the one observed in Ref. [\citenum{Krull2017}]. Once the collective mode of a system exceeds the system's smallest possible Higgs-mode $\omega_H$, this mode couples to $\omega_H$ and is therefore pushed below the quasiparticle continuum. The effect is not dominant on the border to the s-wave state, since the order parameter is fully gapped on this site.\\

\bibliography{marvinsis1}

\end{document}